\documentclass[aps,twocolumn,showpacs,floatfix]{revtex4}

\usepackage{graphicx}
\usepackage{dcolumn}
\usepackage{bm}
\usepackage{epsf}
\usepackage{subfigure}
\usepackage{epstopdf}
\usepackage{amsmath}
\usepackage{amssymb}

\newcommand{\e}{{\rm e}}

\begin{document}

\title{Extracting chemical energy by growing disorder: Efficiency at maximum power}

\author{Massimiliano Esposito}
\altaffiliation[]{Also at Center for Nonlinear Phenomena and Complex Systems,
Universit\'e Libre de Bruxelles, Code Postal 231, Campus Plaine, B-1050 Brussels, Belgium.\\}
\author{Katja Lindenberg}
\affiliation{Department of Chemistry and Biochemistry and BioCircuits
Institute, 
University of California, San Diego, La Jolla, CA 92093-0340, USA}
\author{Christian Van den Broeck}
\affiliation{Hasselt University, B-3590 Diepenbeek, Belgium}



\begin{abstract}
We consider the efficiency of chemical energy extraction from the environment by the 
growth of a copolymer made of two constituent units in the entropy-driven regime.
We show that the thermodynamic nonlinearity associated with the information processing aspect
is responsible for a branching of the system properties such as power,
speed of growth, entropy 
production, and efficiency, with varying affinity. The standard linear
thermodynamics 
argument which predicts an efficiency of $1/2$ at maximum power is
inappropriate because the regime 
of maximum power is located either outside of the linear regime or on a separate
bifurcated branch, and because the usual thermodynamic force is not the natural variable for this optimization.
\end{abstract}


\maketitle

\section{Introduction}

Carnot efficiency is one of the cornerstones of thermodynamics since it 
leads to the definition of entropy and the Second Law of thermodynamics. 
It expresses a fundamental limitation on how much work can be extracted
from a heat flow. A less studied but arguably more relevant question for many
isothermal chemical and biological processes is how much chemical energy a 
system can extract from its environment
by increasing the system's configurational entropy. 
Thermodynamics does, in fact, also prescribe a limit, even though at 
first sight it appears to be almost trivial: the energy extracted
by such an isothermal transfer can be carried out with $100\%$ efficiency.
However, there is a crucial additional condition, namely, that this efficiency
can only be reached -- just as in the case of Carnot efficiency -- by a
reversible, infinitely slow process. Hence $100\%$ efficiency is achieved 
for a process with zero power output. The question of efficiencies at finite
power should thus be addressed.

In the context of thermal machines, a straightforward analysis based on \emph{linear} 
irreversible thermodynamics teaches us that, as one moves away from the reversible regime, 
the power goes through a unique maximum, and that the efficiency at this maximum is, 
at most, $50\%$ \cite{VandenBroeck05,EspLindBroeckPRL}. The same argument can easily be  
extended to the transformation between different forms of chemical energy.
However, the above prediction may not apply for several reasons. First, the point 
of maximum power does not necessarily lie in the linear regime.
Second, thermodynamic nonlinear effects can give rise to bifurcated branches. 
Finally,  the above-mentioned efficiency is attained upon maximization
with respect to the thermodynamic force associated with the power generating flux. While this is
a natural set-up in many problems, it may not always correspond to the relevant scenario.

In this paper, we investigate the efficiency of a chemical entropy-driven
process of capital importance in biophysics, namely, copolymer synthesis
\cite{Bennett79,08AndrieuxGaspardPNAS,09AndrieuxGaspardJCP,JarzynskiPNAS}, 
see also \cite{Volkenstein,Smith}. As exemplified by the copolymer DNA, 
guardian of genetic information, such processes are essential for biological 
information processing. We will show that the above-mentioned complications 
are present in this generic model. In particular, the thermodynamic nonlinearity 
associated with the information processing aspect is responsible for a branching 
of the systems properties such as power, speed of growth, efficiency and
entropy production, 
as one varies the affinity. Furthermore, the regime of maximum power is located
either outside of
the linear regime or on the separate bifurcated branch. Finally, it turns out that the
thermodynamic force is not a natural control variable in the present model.
 While our (exact)
analysis is carried out for the simplest possible model, namely, copolymer
synthesis with two constituent building blocks, our findings suggest that
chemical information processing usually operates in the far-from-equilibrium
regime, with unique features due to the entropic contribution.

In Sec.~\ref{thermodynamics} we present the basic thermodynamic formulas that
define our system.  In
Sec.~\ref{kinetics} we present the detailed kinetic description of our model,
whose results are discussed in detail in Sec.~\ref{effmax}.  In particular,
it is here that we exhibit the correct and unexpected results for the
efficiency at maximum power, results that arise entirely from the nonlinear
nature of the problem. A brief recapitulation is presented in
Sec.~\ref{conclusions}.

\section{Thermodynamics}
\label{thermodynamics}

We begin with some well-known relations for isothermal systems.
Consider a spontaneous chemical process involving particles of different types
labeled by $j$, with corresponding particle number $N_j$ and chemical potential $\mu_j$.
The system is in thermal and mechanical equilibrium at temperature $T$ and pressure $P$. 
The total Gibbs free energy
\begin{equation}
G=U+PV-TS=H-TS=\sum_i \mu_i N_i
\end{equation}
evolves toward a minimum value, so that $dG \leq 0$.  
Alternatively, to characterize the evolution of the isothermal system we write
\begin{eqnarray}
\label{def1}
d S &=& d_i S + d_e S \label{def2} \nonumber\\
 && T d_i S = -\sum_j \mu_j d_iN_j \nonumber \\
&&T d_e S = dH - \sum_j \mu_j d_eN_j.
\end{eqnarray}
We have separated the total entropy change into two contributions. The first one, $d_iS$, 
is the always-positive part of the entropy change, called the internal
entropy production. The other is the contribution to the entropy change due to 
exchange processes between the system and its environment, and can be positive or negative.  
 Associated with these 
contributions, we have written the change in he number of particles of type $j$ as
\begin{equation}
dN_j = d_eN_j +d_iN_j,
\end{equation}
where the first contribution is due to exchange of particles with the
environment, and the second is the internal change caused by the chemical reaction.
We take the system to be closed, i.e., it exchanges only energy but not particles with 
the environment, so that $d_eN=0$. These definitions lead to consistency between the 
statements that the system evolves toward a minimum in the Gibbs free energy and that 
the internal entropy production of this chemical system has to be positive~\cite{Prigogine}, that is, 
\begin{equation}
d_i S = dS -\frac{dH}{T} =-\frac{\sum_j \mu_j  dN_j}{T} = - \frac{dG}{T}\geq 0 .\label{ep}
\end{equation}
Obviously, for a reversible transformation with zero internal entropy
production $d_iS=0$. 

We now turn to the simplest scenario of copolymer synthesis. The system
consists of a bulk-phase containing two types of monomer units, $1$ and
$2$, which can attach or detach at the endpoint of a single linear copolymer. 
We identify the four states $1f$, $1c$, $2f$ and $2c$. Here $jf$ represent free bulk
monomers and $jc$ represent monomers attached to the copolymer.
Since the number of each type of monomer is conserved, one has $d N_{1f}=-d N_{1c}$ and $d N_{2f}=-dN_{2c}$. 
The entropy production (\ref{ep}) can thus be written in the familiar bilinear form 
\begin{eqnarray}
\dot{S}_i &\equiv&  \frac{d_iS}{dt} = (\mu_{1f}-\mu_{1c})\frac{dN_{1c}}{dt} +
(\mu_{2f}-\mu_{2c})\frac{dN_{2c}}{dt} \nonumber\\
&=& A_1 J_1 + A_2 J_2,
\end{eqnarray}
with the affinities $A_{j}=(\mu_{jf}-\mu_{jc})$ and the conjugate fluxes
$J_{j}= d N_{jc}/dt$. 

In view of the relation $d_i S= d S - dH/T$, we rewrite the entropy production as
\begin{equation}
\dot{S}_i = \big(s_1 - \frac{h_1}{T} \big) J_1 + \big(s_2 - \frac{h_2}{T} \big) J_2 + D (J_1+J_2).
\end{equation}
Here $h_j$ is the change of enthalpy per monomer upon transfer from the 
bulk to the copolymer. The crucial point, which has been discussed in detail in the literature  \cite{Bennett79,08AndrieuxGaspardPNAS,09AndrieuxGaspardJCP,JarzynskiPNAS}, is to realize
that the average change of entropy upon transfer of a monomer 
from the bulk to the copolymer contains two contributions. 
One is the monomer entropy, $s_j$,  due to the change in the monomer degrees 
of freedom and in the monomer internal structure between 
the free monomer in solution and the monomer inside the copolymer.
The other is the configurational entropy denoted by $D$, due to the change in the
information 
stored in the polymer sequence that occurrs when a monomer is added to the
copolymer. 
It is given by the Shannon entropy 
\begin{equation}
D = - \lim_{l \to \infty} \frac{1}{l} \sum_{\omega} P_{\omega} \ln P_{\omega},
\end{equation}
where $l$ is the copolymer length in monomer units and $P_{\omega}$ is the
probability of a copolymer with monomer sequence $\omega$. In the absence of
correlations, the Shannon entropy is expressed solely in terms of the monomer
abundance probabilities $p_1=p$ and $p_2=1-p$,
\begin{eqnarray}
D=-p \ln p - (1-p) \ln (1-p) \label{disorder}.
\end{eqnarray}
For simplicity, we further assume that monomer entropy and enthalpy changes upon
transfer of a monomer 
from the bulk to the copolymer and vice versa have the same value for both
monomers, that is,
\begin{eqnarray}
\epsilon \equiv \frac{h_1}{T}- s_1 = \frac{h_2}{T}- s_2.
\end{eqnarray}
We henceforth call $T\epsilon$ the monomer ``free enthalpy.''
Introducing the net speed of growth of the copolymer, $v=J_1+J_2$, the 
entropy production can finally be written as
\begin{eqnarray}
\dot{S}_i = A v \geq 0, \label{EntropyProdmodel}
\end{eqnarray}
where the total affinity is given by
\begin{eqnarray}
A= D - \epsilon . \label{Affinities}
\end{eqnarray}
The expression~(\ref{EntropyProdmodel}) for the entropy production in the 
steady state regime of the growing copolymer has been derived in Refs.
\cite{Bennett79,08AndrieuxGaspardPNAS,JarzynskiPNAS,09AndrieuxGaspardJCP}.
It is interesting to realize that the affinity is not an obvious control parameter 
due to its dependence on $D$ which is in turn a nontrivial function of $\epsilon$. 
Only $\epsilon$ can be easily  controlled externally by changing the concentration of the monomers in solution.

The power at which $\epsilon$, the free enthalpy divided by temperature,
is extracted from the surroundings by copolymer growth is given by
\begin{eqnarray}
\mathbb{P} =  \epsilon v= (D-A) v \label{powerGen}.
\end{eqnarray}
The efficiency $\eta$ of the process is defined as the ratio of this 
power over the cost $D v$ of the entropy growth per unit time,
\begin{eqnarray}
\eta = \frac{\epsilon v}{D v}= \frac{D-A}{D} \label{EffGen}.
\end{eqnarray}

In the reversible limit with $A , v \to 0$, the efficiency of the process
becomes optimal, $\eta=1$, but the extracted power goes to zero. The standard
prediction from linear thermodynamics that arrives at an efficiency of $50\%$ at
maximum power is obtained upon expanding the velocity in terms of the
affinity, $v=L A$, with $L$ the linear response coefficient. Within this 
approximation the power becomes $\mathbb{P} \approx L (D-A) A$. Note that this power attains its
maximum for $A=D/2$ with the corresponding efficiency $\eta=1/2$, if we assume that $D$ is kept
constant. However, below we will investigate the more natural optimization with respect of
$\epsilon$, since this is the natural and easily controllable variable related to the free enthalpy flux.
Whatever control variable is used, we will see in Sec. \ref{effmax} that the true maximum is beyond the
reach of this linear expansion (and even of a nonlinear continuation of this
expansion).


\section{Kinetics}\label{kinetics}

We now turn to the detailed kinetic description of the copolymerization process, 
which will allow us to identify the expressions for $v$ and $p$ in the context
of a full nonlinear analysis. Let us call $k_{+j}$ and $k_{-j}$ the rates of
insertion and removal, respectively, of monomer  $j=1,2$. 
Because the free enthalpy of the monomers has been assumed to be the same, 
the ratios of the reaction rates are given by
\begin{eqnarray}
\frac{k_{+1}}{k_{-1}}=\frac{k_{+2}}{k_{-2}}=\e^{-\epsilon} .\label{def_epsilon}
\end{eqnarray}
The  fraction $p$ of monomers of type ${1}$ present in the copolymer  in the 
regime of steady growth can be determined by the following self-consistency
argument. The ratio $p/(1-p)$ of the number of ${1}$ versus ${2}$ monomers
in an ensemble of copolymers has to be equal to the ratio of their net rates of
attachment. For monomer  ${1}$, this net rate is the pure rate of attachment,
$k_{+1}$, minus the rate of detachment, which is $-k_{-1}p$. The factor $p$
arises from the fact that detachment is only possible when the monomer at the
tip  of the copolymer is of type ${1}$, and this occurs with probability
$p$. The net rate of attachment for ${2}$ is similarly given by $k_{+2} -
k_{-2}(1-p)$. We thus conclude that 
\begin{equation}
\frac{k_{+1}-k_{-1}p}{k_{+2}-k_{-2}(1-p)}=\frac{p}{1-p}.\label{ma}
\end{equation}
The solution of the resulting quadratic equation for $p$ reads
\begin{eqnarray}
p = \frac{a-\sqrt{a^2-4(k_{-1}-k_{-2})k_{+1}}}{2(k_{-1}-k_{-2})} , \label{p}
\end{eqnarray}
where $a=k_{+1}+k_{+2}+k_{-1}-k_{-2}$.
By a similar argument we find that the speed of growth of the copolymer, given
by the rate of attachment $k_{+1}+k_{+2}$ minus the rate of detachment
$k_{-1}p+k_{-2}(1-p)$, is given by
\begin{eqnarray}
v= k_{+1}-k_{-1}p+k_{+2}-k_{-2}(1-p) \label{speed}.
\end{eqnarray}

We note from Eqs.~(\ref{disorder}), (\ref{EntropyProdmodel}), (\ref{Affinities}), (\ref{p}) and (\ref{speed}) 
that equilibrium, $v=0$ and $A=0$, occurs at $\epsilon=\ln 2$ with $p=1/2$ and $D=\ln 2$. 
For smaller (larger) values of $\epsilon$, $A>0$ ($A<0$) and the copolymer
is synthesized (degraded), i.e., $v > 0$ ($v < 0$). 
Of specific interest to us is the surprising regime of
entropy-driven growth, $A>0$  and $v > 0$, but with $\epsilon>0$ 
\cite{Bennett79,08AndrieuxGaspardPNAS,09AndrieuxGaspardJCP,JarzynskiPNAS}. 
Under the simplifications assumed in our model, this occurs when $0 \leq \epsilon \leq
\ln 2$. Monomers are pumped uphill against the free enthalpy barrier $\epsilon \geq 0$
under the influence of the entropic contribution $D$ to the affinity. The power
$\mathbb{P}$ (enthalpy per unit time) extracted from the copolymerization
dynamics is positive in this entropy driven regime, cf. Eq.~(\ref{powerGen}),
with corresponding efficiency  given in Eq.~(\ref{EffGen}).

\section{Efficiency at maximum power}\label{effmax}

\begin{figure}
\centering 
\begin{tabular}{c@{\hspace{0.5cm}}c}
\rotatebox{0}{\scalebox{0.4}{\includegraphics{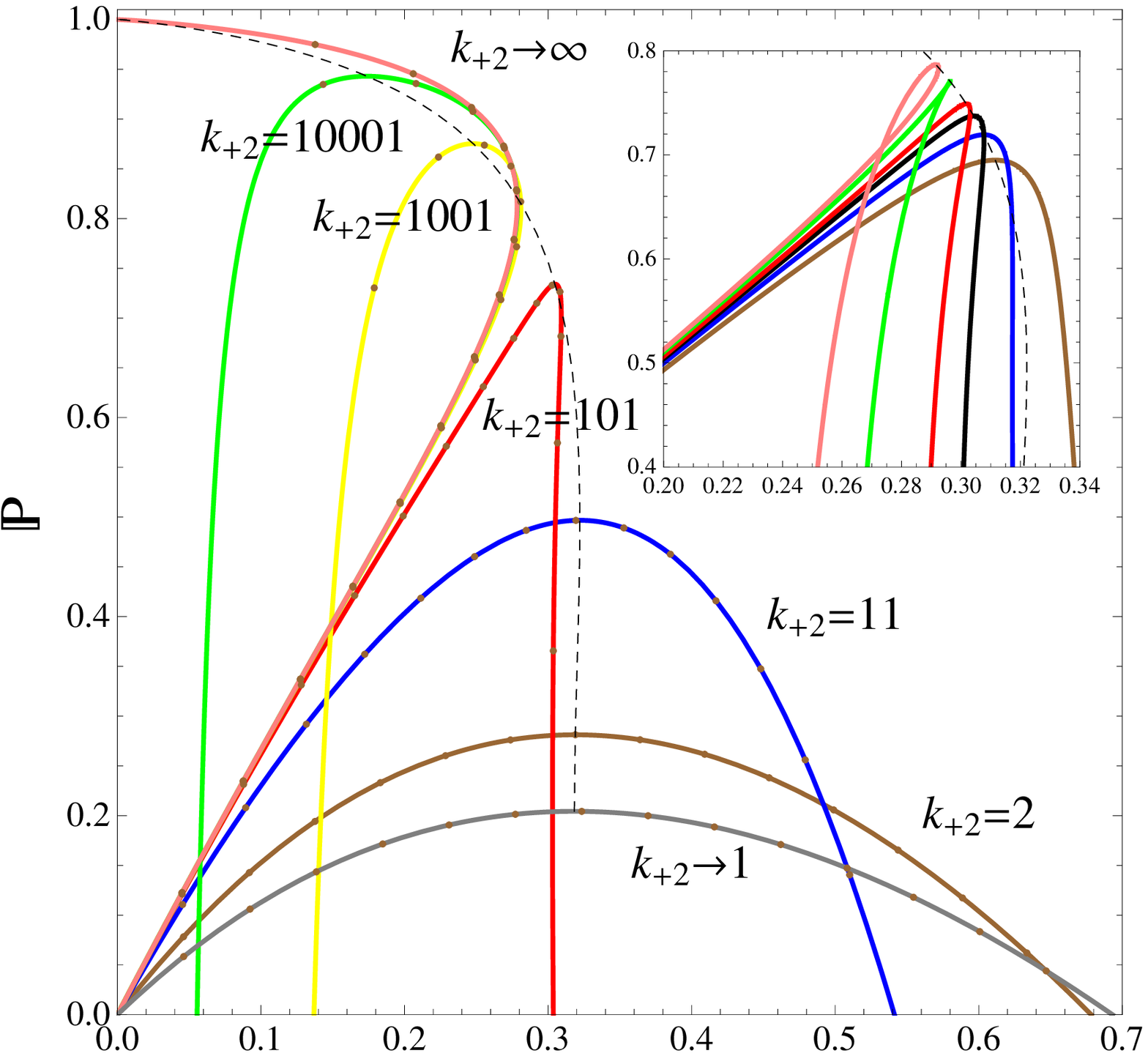}}} \vspace{-0.38cm} \\
\hspace{-0.35cm}
\rotatebox{0}{\scalebox{0.415}{\includegraphics{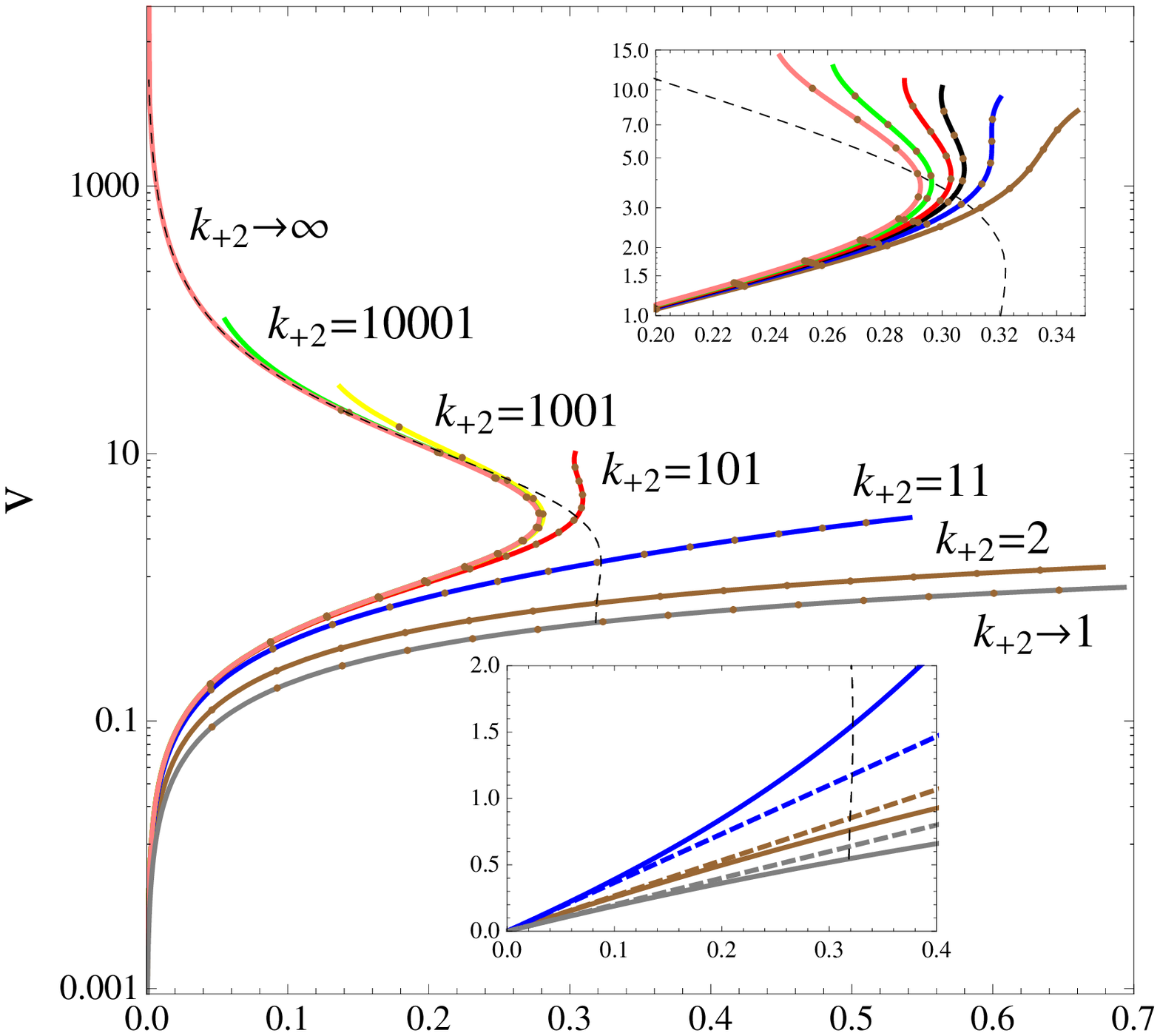}}} \vspace{-0.5cm}\\
\hspace{-0.25cm}
\rotatebox{0}{\scalebox{0.402}{\includegraphics{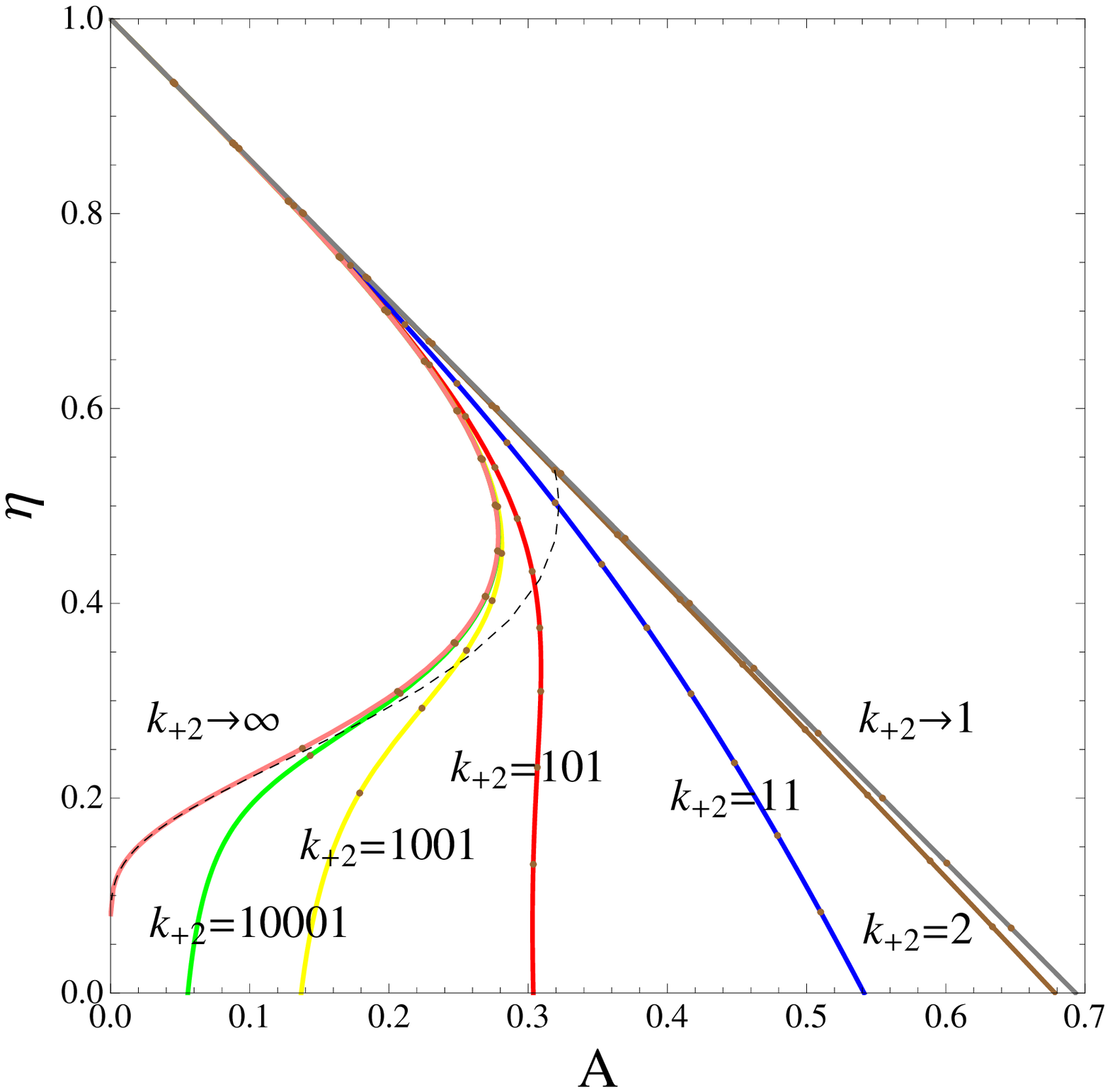}}} 
\end{tabular}
\caption{(Color online) 
The full thick curves represent the parametric dependence on $0 \leq \epsilon
\leq \ln 2$ of the power $\mathbb{P}$, the copolymerization speed $v$, and the
efficiency $\eta$ on the ordinate and the affinity $A$ as abscissa. The point
$\epsilon=\ln 2$ is located at the origin of the axes for $\mathbb{P}$ and
$v$ and at $\eta=1$ and $A=0$ for $\eta$. The small dots
along the curves are separated by $\Delta \epsilon = (\ln 2)/14$ to indicate how
fast $\epsilon$
changes along the curves. Different thick curves correspond to different choices
of $k_{+2}$, with $k_{+1} \leq k_{+2} \leq \infty$. Without loss of generality
we set $k_{+1}=1$ (time rescaling). The thin dashed curves
intersect the thick curves where the value of $\epsilon$ corresponds to
maximum power with respect to $\epsilon$. The curves in the inset in the 
$\mathbb{P}$ plot and in the upper inset 
of the $v$ plot correspond to $k_{+2}=65$, $85$, $105$, $121$, $160$, and $200$. 
The dashed curves in the lower inset of this plot represent the linear response
predictions $v=LA$ for $k_{+2}=1,2,11$.}
\label{plot1}
\end{figure}

\begin{figure}[h]
\centering 
\rotatebox{0}{\scalebox{0.4}{\includegraphics{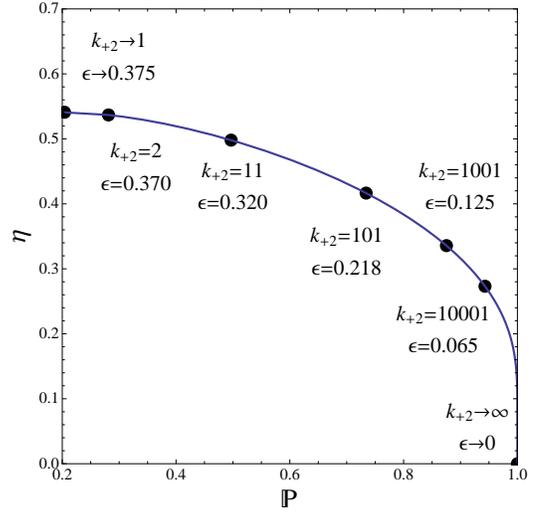}}}
\caption{(Color online) 
Efficiency $\eta$ and power $\mathbb{P}$ corresponding to the maximum power
denoted simply by $\epsilon$ in the figure for different values of $k_{+2}$.
We have set $k_{+1}=1$.}
\label{plot2}
\end{figure}

\begin{figure}[h]
\centering 
\rotatebox{0}{\scalebox{0.4}{\includegraphics{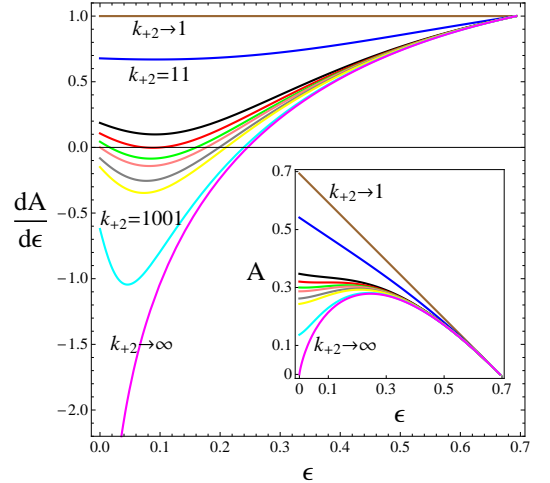}}}
\caption{(Color online) Derivative of the affinity with respect to $\epsilon$
and affinity (in the inset) as a function of $\epsilon$. The different curves
correspond to $k_{+2}=0,11$, $65$, $85$, $105$, $121$, $160$, $200$,
$1001$, and $\infty$ and $k_{+1}=1$.}
\label{plot3}
\end{figure}

\begin{figure}[h]
\centering 
\rotatebox{0}{\scalebox{0.4}{\includegraphics{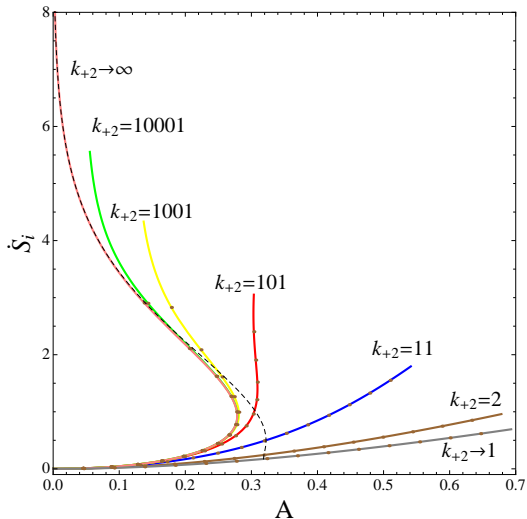}}}
\caption{(Color online) 
Same type of plot as Fig. \ref{plot1} but for entropy production.}
\label{plot4}
\end{figure}

To study the main question of interest, namely, the regime of maximum power and
its corresponding efficiency, we choose convenient variables. We
note that the model is described by four kinetic constants, but the latter are not
independent since they obey the relation  Eq.~(\ref{def_epsilon}). Furthermore,
one of them can be set equal to $1$ by an appropriate choice of the time unit,
e.g., $k_{+1}=1$. As the remaining two degrees of freedom, we choose $\epsilon$
and $k_{+2}$. We then have explicit functional expressions for all the
other quantities $k_{-1}=e^{\epsilon}$, $k_{-2}=e^{\epsilon}k_{+2}$, $p$ and
$v$, cf. Eqs.~(\ref{p}) and (\ref{speed}), and hence also $D$, $A$, $\mathbb{P}$, $\dot{S}_i$,
and $\eta$, see
Eqs.~(\ref{Affinities}), (\ref{powerGen}), (\ref{EffGen}), (\ref{disorder}) and
(\ref{EntropyProdmodel}).
Other relations between, for example, $\mathbb{P}$ and $A$, can then be obtained
by parametric elimination. The quantities $\mathbb{P}$, $v$, $\eta$, $A$, and $\dot{S}_i$
can easily be calculated numerically. The results are summarized in Figs.
\ref{plot1}, \ref{plot2}, \ref{plot3}, and \ref{plot4}. We
next turn to a discussion of these figures, 
supplemented with corresponding analytic calculations.

The information contained in Fig.\ref{plot1} is detailed in the caption.
The most striking feature in this figure is the existence of two 
different branches for the power and velocity in terms of the affinity. 
The transition between the two branches occurs when
\begin{eqnarray}
\frac{d (\mathbb{P},v)}{d A} 
= \frac{d (\mathbb{P},v)}{d \epsilon} \bigg(\frac{d A}{d \epsilon}\bigg)^{-1}
\end{eqnarray} 
diverges. $(\mathbb{P},v)$ indicates $\mathbb{P}$ or $v$. 
Since $\partial (\mathbb{P},v)/\partial\epsilon$ is an analytic function of
$\epsilon$, the new branch appears when $\partial A/\partial\epsilon$ touches
zero. As long as the latter quantity remains positive, which is the case for 
$k_{+2}$ smaller than a certain critical value, cf. Fig.~\ref{plot3}, the power 
and velocity can be seen as a true function of $A$. Branching takes place at the critical 
point, characterized by $\partial A/\partial\epsilon=\partial^{2}A/\partial\epsilon^2=0$, 
resulting in $k_{+2} \approx 84.33$ and $\epsilon=0.088$, see again Fig. \ref{plot3}. 
For values of $k_{+2}$ larger than this critical value, power is 
no longer a proper function of $A$, as two branches appear, with  two values of
$(\mathbb{P},v)$ for a given value of $A$. While along the linear branch and its continuation 
the affinity decreases with $\epsilon$, the affinity \emph{increases} with
$\epsilon$ on
the new lower branch, cf. the inset in Fig.~\ref{plot3}. This remarkable result
implies that we can approach low values of affinities via a nonlinear branch which 
is distinct from the branch predicted by linear response theory and its continuation.  
We note that the entropy production itself becomes a bi-valued 
function in terms of the affinity, as can be seen in Fig.~\ref{plot4}.
Naively, one would expect that entropy production and affinity both provide consistent 
measures for the distance from equilibrium. This is clearly not the case in the present 
model, where the entropy production is a decreasing function of the affinity on
the upper 
nonlinear branch. In particular, for very large values of $k_{+2}$ one finds
that the 
entropy production becomes very large while the affinity goes to zero.  
We conclude that the affinity is not a reliable measure for the distance from equilibrium. 

To explore the region close to equilibrium and,  in particular, the linear
response regime, we write $\epsilon=\ln2 -\delta \epsilon$ and expand 
in powers of $\delta \epsilon$. From Eq.~(\ref{disorder}) with Eqs.~(\ref{p}) and
(\ref{def_epsilon}) we find
\begin{eqnarray}
D= \ln2-\alpha \; \delta \epsilon^2 + {\cal O}(\delta \epsilon^3),
\label{EntrLinResp}
\end{eqnarray}
where 
\begin{equation}
\alpha = \frac {(k_{+1}-k_{+2})^2}{2(k_{+1}+k_{+2})^2}.
\end{equation}
For the affinity, we find from Eq.~(\ref{Affinities}) that
\begin{eqnarray}
A = \delta \epsilon -\alpha \delta \epsilon^2 + {\cal O}(\delta \epsilon^3). \label{AffLinResp}
\end{eqnarray}
The efficiency thus becomes
\begin{eqnarray}
\eta &=& 1-\frac{\delta \epsilon}{\ln 2} + 
\frac{\alpha}{\ln 2} \delta \epsilon^2 + {\cal O}(\delta \epsilon^3) \nonumber\\
&=& 1-\frac{A}{\ln 2} + {\cal O}(\delta \epsilon^3).
\end{eqnarray}
This linear dependence of the efficiency on the affinity close to equilibrium is clearly identified 
in the upper left region of the third affinity plot in Fig.~\ref{plot1}, while
the
corresponding behavior of the affinity in terms of $\epsilon$, cf.
Eq.~(\ref{AffLinResp}), is observed
in the lower left region of the inset of Fig.~\ref{plot3}. 
In this regime close to equilibrium we find the standard linear response
relations
\begin{eqnarray}
v &=&  L A + {\cal O}(\delta \epsilon^2) \label{LinRespForV} \\
\mathbb{P} &=& \ln 2 \; L A + {\cal O}(\delta \epsilon^2), \label{LinRespForP}
\end{eqnarray}
with the Onsager coefficient given by $L=4 k_{+1} k_{+2}/(k_{+1}+k_{+2})$, cf.
the lower left regions of the power and speed plots in Fig.~\ref{plot1}. 
Note also that the Onsager coefficient becomes independent of $k_{+2}$ 
in the limit $k_{+1} \ll k_{+2}$, where $L=4 k_{+1}$.
 
We have seen that linear response predicts an efficiency at maximum power of $50\%$.
However, as announced earlier, this result is not correct. This is seen in 
Fig.~\ref{plot1} or in 
Fig.~\ref{plot2}, where the affinity is clearly above the value $1/2$ in the
regime 
``closest" to equilibrium. The explanation is that maximum power
occurs beyond the reach of linear response theory, as can clearly be seen in the
lower inset of 
the $v$ plot in Fig.~\ref{plot1}, where the linear response curves (dashed lines) 
become inaccurate at maximum power. 
Furthermore, we note that the point of maximum power moves onto the 
nonlinear branch as $k_{+2}$ grows, now occurring at decreasing values of $A$. 
So, even though we are approaching a regime of low power output with decreasing 
affinity, we do so via the nonlinear branch, where the prediction of linear 
response theory utterly fails.
The main conclusion is that, while there is indeed a regime of linear
response, it is unable to describe the region of maximum power, which 
always occurs outside the regime of validity of the linear law.

To complete our analysis, we explore in detail the limiting cases $k_{+2} \to
k_{+1}$ and $k_{+2} /k_{+1} \to \infty$. For transparency, we explicitly
retain $k_{+1}$ instead of setting it equal to unity.
In the limit where $k_{+2} \to k_{+1}$, we find that 
\begin{equation}
p=\frac{1}{2}, \qquad  v=k_{+1}(2-\e^{\epsilon}), \qquad D=\ln 2.
\end{equation}
This
leads to an efficiency $\eta=\epsilon/\ln 2=1-A/\ln 2$, as observed in 
Fig.~\ref{plot1}. In this limit, the value of $\epsilon$ leading to maximum
power is obtained as the solution of the transcendental equation $2
\e^{-\epsilon}-\epsilon=1$, namely, $\epsilon \approx 0.375$. At maximum power
we
thus get 
\begin{equation}
\mathbb{P}\approx 0.204 k_{+1} \mbox{  and   } \eta\approx 0.541,
\end{equation}
as seen in
Figs.~\ref{plot1} and \ref{plot2}. As an immediate consequence, we also find
$v\approx0.545 k_{+1}$ and $A\approx0.318$, as observed in Fig.~\ref{plot1}. 

In the limit $k_{+2} \to \infty$, where 
\begin{eqnarray}
&& p=1-\e^{-\epsilon}, \ \ v=k_{+1}
\e^{\epsilon}\frac{(2\e^{-\epsilon}-1)}{(\e^{-\epsilon}-1)}, \nonumber \\
&& D=\e^{-\epsilon}\epsilon-(1-\e^{-\epsilon}) \ln (1-\e^{-\epsilon}),
\end{eqnarray}
the
efficiency reads
$\eta=-\epsilon/[\e^{-\epsilon}\epsilon+(1-\e^{-\epsilon}) \ln
(1-\e^{-\epsilon})]$. The numerical results of Fig.~\ref{plot1} suggest that
maximum power in this limit occurs for $\epsilon$ very close
to zero. We therefore expand the velocity around $\epsilon=0$ and find
$v=-\sqrt{k_{+1} k_{+2}}+(k_{+1}+k_{+2}) \epsilon/2 + {\cal O}(\epsilon^2)$.
Using Eq.~(\ref{powerGen}), we find that maximum power occurs at
$\epsilon=-\sqrt{k_{+1} k_{+2}}/(k_{+1}+k_{+2})$, resulting in
$\mathbb{P}=\sqrt{k_{+1}}k_{+2}/(k_{+1}+k_{+2})$. For $k_{+2}\to \infty$ the
latter becomes 
\begin{equation}
\mathbb{P}=1,
\end{equation}
as observed in Figs.~\ref{plot1} and \ref{plot2}.
Similarly, by expanding $\eta$ to first order around $\epsilon=0$ and using the
value we found for $\epsilon$ at maximum power, we find that
\begin{equation}
 \eta \to 0,
\end{equation}
as observed in Figs.~\ref{plot1} and \ref{plot2}.

\section{Conclusions}
\label{conclusions}

Using a simple model of copolymerization, we have shown that free enthalpy can 
be extracted from the environment in response to the entropic force corresponding 
to the information stored in a growing copolymer sequence. 
The thermodynamic nonlinearity associated with the information processing aspect 
is responsible for a branching of the dependence on the affinity of system
properties such as power, speed of growth
and efficiency. The nonlinear regime occuring 
after the branching is particularly surprising since the entropy production keeps
increasing even as the affinity begins to decrease.
We identified a regime of linear response where the efficiency of the energy 
extraction is optimal (equal to $1$), but where, as usual, the power output goes
to zero. 
Considering instead the efficiency at maximum power, we found that the universal 
prediction of linear response theory (efficiency equal to $1/2$) is inappropriate for this model. 
The reason is that the copolymerization generating maximum power occurs 
far from equilibrium in a region not accessible to linear response theory. 
Our results suggest a possible self-powering mechanism for nonequilibrium systems that can 
extract chemical energy from their surroundings by growing their internal
structural information. 


\section*{Acknowledgments}

M. E. is supported by the FNRS Belgium (charg\'e de recherches) and 
by the government of Luxembourg (Bourse de formation recherches). This work was
partially supported by the National Science Foundation under grant No.
PHY-0354937.



\end{document}